\documentclass[lettersize,journal,onecolumn]{IEEEtran}
\usepackage{amsmath}
\usepackage{algpseudocode}
\usepackage{algorithm}
\usepackage[cmintegrals]{newtxmath}
\usepackage{cite}
\usepackage[utf8]{inputenc}
\usepackage[english]{babel}
\usepackage[]{verbatim}
\usepackage{syntonly}
\usepackage{textcomp}
\usepackage{upgreek}
\usepackage{graphicx} 
\usepackage{varioref}
\usepackage[noabbrev]{cleveref}
\usepackage{textcomp,xcolor}
\usepackage{acronym}
\usepackage{balance}
\newtheorem{definition}{Definition}
\usepackage{balance}
\usepackage[normalem]{ulem}
\usepackage{xcolor,cite,etoolbox}

\acrodef{HAPS}{high-altitude platform station}
\acrodef{RIS}{reconfigurable intelligent surfaces}
\acrodef{IoT}{Internet-of-Things}
\acrodef{BS}{base station}
\acrodef{UE}{user equipment device}
\acrodef{CS}{control station}
\acrodef{TN}{terrestrial network}
\acrodef{NTN}{non-terrestrial network}
\acrodef{RF}{radio frequency}
\acrodef{QoS}{quality-of-service}
\acrodef{SNR}{signal-to-noise ratio}
\acrodef{OFDM}{orthogonal frequency division multiplexing}
\acrodef{RE}{resource-efficiency}
\acrodef{MINLP}{mixed-integer nonlinear program}

\makeatletter 
\pretocmd\@bibitem{\color{black}\csname keycolor#1\endcsname}{}{\fail}
\newcommand\citecolor[1]{\@namedef{keycolor#1}{\color{black}}}
\makeatother

\citecolor{9709573}
\citecolor{bjornson2022reconfigurable}
\citecolor{rivera2022optimization}

\begin{document}

	\title{Resource-Efficient HAPS-RIS Enabled Beyond-Cell Communications}
	\author{Safwan Alfattani,~\IEEEmembership{Member,~IEEE,} Animesh Yadav,~\IEEEmembership{Senior Member,~IEEE,} Halim Yanikomeroglu,~\IEEEmembership{Fellow,~IEEE,} Abbas Yonga{\c{c}}oglu,~\IEEEmembership{Member,~IEEE}
\thanks{This research has been supported in part by  King AbdulAziz University, Saudi Arabia, and in part by Huawei Canada Company Ltd.

S. Alfattani is with King AbdulAziz University, Saudi Arabia  (e-mail: smalfattani@kau.edu.sa). A. Yadav is with the Department of ECET, Minnesota State University, Mankato, MN, 56001, USA (e-mail: animesh.yadav@mnsu.edu). H. Yanikomeroglu is with the Department of Systems and Computer Engineering, Carleton University, Ottawa, ON, K1S 5B6, Canada (e-mail: halim@sce.carleton.ca). A. Yongacoglu is with the University of Ottawa, ON, Canada (e-mail: yongac@uottawa.ca).   }
}
	\maketitle

\begin{abstract}
 In the future, urban regions will encounter 
a massive number of capacity-hungry
devices. Relying solely
on terrestrial networks for serving all UEs will be
a cost-ineffective approach. Consequently,
with the anticipated supply and demand mismatch,
 several UEs will be unsupported.
 To offer service to the left-out UEs, we employ an energy-efficient
and cost-effective \emph{beyond-cell communications} approach,
which uses reconfigurable intelligent surfaces (RIS) on
a high-altitude platform station (HAPS). Particularly,
unsupported UEs will be connected to a dedicated control
station (CS) through RIS-mounted HAPS. A novel resource-efficient
optimization problem is formulated that maximizes the
number of connected UEs, while minimizing the
total power consumed by the CS
and  RIS. Since the resulting
problem is a mixed-integer nonlinear program (MINLP), a low-complexity two-stage algorithm is
developed. Numerical results demonstrate that the proposed algorithm
outperforms the benchmark approach in terms of the percentage
of connected UEs and the resource-efficiency (RE). Also,
the results show that the number of connected UEs
 is more sensitive to transmit power at the CS than the HAPS size.
\end{abstract}
	

\begin{IEEEkeywords}
\ac{HAPS}, \ac{RIS}, \ac{NTN}, \ac{QoS}
\end{IEEEkeywords}	
\acresetall
	\section{Introduction}
	Future wireless networks are expected to face unprecedented demands for continuous and ubiquitous connectivity due to the increasing number of mobile users, the evolving deployment of \ac{IoT} networks, and the growing number of novel use cases. Increasing \ac{BS} density and using relays may be a straightforward solution to cope with this unforeseen situation, but it comes at the cost of high capital and operational expenditures. Alternatively, a green and an energy-efficient solution that utilizes reconfigurable
	intelligent surfaces (RIS) in terrestrial networks (TNs) has been introduced \cite{huang2019reconfigurable}.
	RIS constitutes a large number of low-cost and nearly passive elements, which can be configured to reflect the incident \ac{RF} signal to a desired direction \cite{di2020smart,huang2019reconfigurable}.
	
	\textcolor{black}{However, the deployment of RIS in terrestrial environments involves several challenges, such as placement inflexibility and channel impairments including excessive path loss and shadowing effects. Instead, in \cite{alfattani2021aerial}, we proposed the integration of RIS 
	on aerial or stratospheric platforms
	and discussed its prospects for wireless systems and services. 
	The main motivations behind this proposition include better channel links, wider coverage and dynamic placement. Moreover, since}
	  energy consumption is a critical issue in aerial platforms, equipping them with a massive number of active antennas would exaggerate the issue. Alternatively, mounting aerial platforms with RIS can provide an energy-efficient solution \cite{ye2022non,shang2022uav,li2022energy,jeon2022energy, aung2022energy}. In addition, due to the favorable wireless channel conditions in non-terrestrial networks (NTN), \ac{RIS}  can support panoramic full-angle reflection serving wider areas with strong line-of-sight (LoS) links \cite{lu2021aerial,ye2022non}. \textcolor{black}{Further, in \cite{alfattani2021link}, we showed that RIS-mounted \ac{HAPS}
	has the potential to outperform 
	RIS-aided terrestrial networks and other RIS-mounted aerial platforms (e.g., UAV)}. \textcolor{black}{Indeed, although the path-loss from a terrestrial node to a UAV is significantly lower than that to a HAPS, the reflection gain offered by HAPS-RIS is much higher, due to the typical large size of  HAPS.}
	
	To reap the benefits of RIS-mounted HAPS (HAPS-RIS), in \cite{alfattani2022beyond}, we proposed a novel \textit{beyond-cell communications} approach. This approach offers service \textcolor{black}{to stranded} terrestrial \acp{UE}, whose either channel conditions are below the required quality-of-service (QoS), or are located in a cell with fully loaded \ac{BS}. In particular, \textcolor{black}{ stranded} UEs get service from a dedicated ground \ac{CS} through the HAPS-RIS. We showed that HAPS-RIS can work in tandem with legacy TNs to support unserved UEs. We also discussed the optimal power and RIS units allocation design schemes to maximize the system throughput and the worst UE rate.

\textcolor{black}{Previous works on RIS-assisted communications consider serving all UEs by optimizing only the RIS phase shifts and the transmit power \cite{huang2019reconfigurable,alfattani2022beyond}.}	
\textcolor{black}{However, due to practical limitations on system resources, including the transmit power of the \ac{CS} and HAPS size (equivalently, the number of RIS units), it might be infeasible to serve all unsupported UEs by CS through HAPS-RIS, especially, for a large set of UEs.}
Accordingly, we formulated a novel optimization problem to maximize the \ac{RE} of the system. 
	\begin{itemize}
	    \item A novel resource-efficient optimization problem that maximizes the percentage of connected \acp{UE} while minimizing the usage of RIS units and transmit power is formulated.  
	    \item Since the resulting problem is a \ac{MINLP} and hard to solve, a low complexity two-stage algorithm is proposed to solve it.
	    \item Through numerical results, we study the impact of HAPS size and QoS requirements of \acp{UE} on the percentage of connected UEs and demonstrate significant improvements in \ac{RE} of the system.
	\end{itemize}
	
	The rest of the letter is organized as follows. In Section \ref{Sec:model},
	the system model is described. Section \ref{Sec:Problem_form} presents the problem formulation. The proposed solution and the algorithm for solving the optimization problem are discussed in Section \ref{Sec:Solution}. Numerical results and discussion are presented in Section \ref{Sec:Results}. Finally, Section \ref{Sec:conclstion} concludes the letter. 
	\begin{figure}[h]
		\centering
		\includegraphics[width=0.85\linewidth]{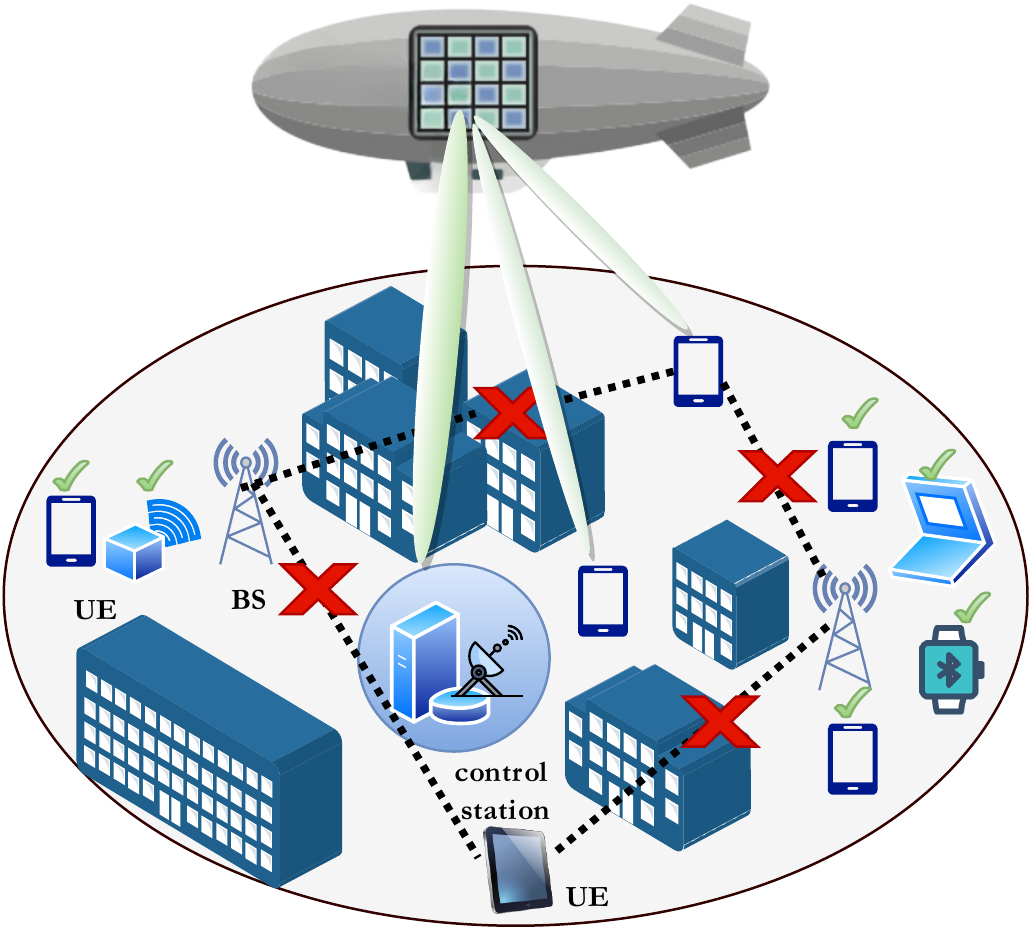}
		\caption{System model for HAPS-RIS assisting beyond cell communications.}
		\label{fig:model}
	\end{figure}
	
	\section{System Model} \label{Sec:model}
    We consider a typical urban region consists of $K$ UEs, $L$ terrestrial \acp{BS}\footnote{\textcolor{black}{BSs have down-tilted antennas dedicated to terrestrial UEs, and can't generally communicate to HAPS.}}, a single HAPS-RIS, and one \ac{CS}\footnote{\textcolor{black}{If the number of unsupported users increases, possible solutions may include increasing the CS transmit power or antenna gain or the number of CS in the HAPS coverage area.}}, as depicted in Fig.~\ref{fig:model}. The UEs are assumed to suffer from severe shadowing and blockages, and NLoS paths, which are typical characteristics of propagation media in urban regions. Based on channel conditions between the \acp{BS} and the UEs, and the maximum serving capacity of the \acp{BS}, a set of $\mathcal{K}_1 = \{1,\ldots,K_1\}$ UEs will be supported by  direct links from the BSs (referred to as \textit{within-cell} communications \cite{alfattani2022beyond}).
	The set of remaining UEs $\mathcal{K}_2 = \{1,\ldots,K_2\}$, which cannot form direct connection with the terrestrial \ac{BS} will be served by the \ac{CS}  via HAPS-RIS (referred to as \textit{beyond-cell} communications \cite{alfattani2022beyond}). The CS is located somewhere in the HAPS coverage area\footnote{\textcolor{black}{According to 3GPP standards \cite{3gpp2017Technical}, a CS (gateway) should be deployed to ensure minimum elevation angle of 5$^\circ$, and the maximum Euclidean distance of 229 km between HAPS and CS} }. Note that $\mathcal{K} = \mathcal{K}_1 \cup \mathcal{K}_2$. We assume that the \ac{CS} serves the stranded UEs in set $\mathcal{K}_2$ using orthogonal subcarriers, and hence, there will be no inter-UE interference. Further, both \textit{within-cell} and \textit{beyond-cell} communications occur in two orthogonal frequency bands, while keeping the subcarrier bandwidth $B_{\rm UE}$ same for both types of communications. As a result, the signals from  \textit{within-cell} UEs will not interfere with the signals from  \textit{beyond-cell} UEs and vice versa.
	
	We assume that \textit{within-cell} UEs are connected optimally with terrestrial  BSs, and hence, our focus in this work is on \textit{beyond-cell} communications. Accordingly, \textcolor{black}{the received signal at UE $k \in \mathcal{K}_2$ on a given subcarrier  can be expressed as
	\begin{equation}
	\label{eq:Rx_signal}
	y_{k }=\sqrt{P^{\rm CS}_{k }} h_{k }  \mathrm{\Phi}_{k} \;x_{{k }}+w_{k},
	\end{equation} where $x_{{k }}$ and $P^{\rm CS}_{k}$ denote the transmitted signal and  power of UE $k$, respectively.  $w_{k}$ denotes} the additive white Gaussian noise (AWGN) $\sim \mathcal{CN}(0,N_0B_{\rm UE})$, where $N_0$ is the noise power spectral density. $h_{k}$ denotes the effective channel gain from the \ac{CS} to the HAPS-RIS and from the HAPS-RIS to UE $k$, and is given by 
	\begin{equation}
	h_{k } = \sqrt{G^{\rm CS} G_{r}^{k} (\textsf{PL}^{\rm{CS-HAPS}-\textit{k}})^{-1}},    
	\end{equation}
	where $G^{\rm CS}$ denotes the 
	antenna gain of  the control station, and $G_{r}^{k}$ is the receiver antenna gain of UE  $k$. $\textsf{PL}^{\text{CS-HAPS}-k} = \textsf{PL}^{\rm{CS}-\rm{HAPS}} \textsf{PL}^{\text{HAPS}-k}$ denotes the effective path loss between the \ac{CS} and UE $k$ via HAPS-RIS. 
	Since the amplitude and phase responses of RIS reflecting units
	are frequency-dependent \cite{bjornson2022reconfigurable} and  each UE uses different subcarrier, \textcolor{black}{each UE is assigned with a distinct and dedicated set of RIS units tuned to its respective subcarrier}. \textcolor{black}{Accordingly, }
	 $\mathrm{\Phi}_{k}$ represents the reflection gain of the RIS corresponding to UE $k$, and is expressed as
	\begin{equation}
	\mathrm{\Phi}_{k}=\sum_{i=1}^{N_{k}} \rho_i e^{-j  \left(\phi_{i,k} - \theta_{i}-\theta_{i,k}\right)},
	\end{equation}
	where $\rho_i$ denotes the reflection loss corresponding to RIS unit $i$. \textcolor{black}{$\theta_{i}$ and $\theta_{i,k}$ represent the corresponding phases between RIS unit $i$ and both the control station and UE $k$, respectively.} $\phi_{i,k}$ represents the adjusted phase shift of RIS unit $i$ \textcolor{black}{corresponding to user $k$}, and $N_{k}$ represents the total number of RIS units allocated to UE $k$. 
	
	Using \eqref{eq:Rx_signal}, the \ac{SNR} at UE $k$ is written as

	\begin{equation}\label{eq:SNR}
	\gamma_{k}= \frac{P_{k}^{\rm CS} \left| h_{k }  \mathrm{\Phi}_{k}\right|^2} { N_0 B_{\rm UE}},
	\end{equation}
	and the corresponding achievable rate can be expressed as
	\begin{equation}\label{eq:R_K2}
	R_{k} = B_{\rm UE} \log_2(1+\gamma_{k}).
	\end{equation}
		
	\section{Problem Formulation}\label{Sec:Problem_form}
	
	\textcolor{black}{Since the required power for the flight and control system of a HAPS is dependent on its type and size, for generalization, we focus on the  power consumption due to communication payload.  Using power efficiently at the HAPS-RIS for activating the RIS units, and
	at the \ac{CS} for transmitting signals offer sustainable \textcolor{black}{network} and longer refueling intervals for HAPS.}
	Also, in case of a large UE density or strict QoS requirement, \textcolor{black}{it might be infeasible} to serve all unserved UEs by the CS via HAPS-RIS. 
	Therefore, we need to select as many UEs that can be supported by the \ac{CS}, while using as minimum system resources (transmit power and RIS units) as possible. Accordingly, we define a novel performance metric known as resource efficiency as follows:
	\begin{definition}
	Resource efficiency (RE) of the beyond-cell communication system, $\eta$, is defined as the ratio between the 
	percentage of served UEs and the average power consumption in dBm by each supported UE, which includes the consumption towards signal transmission and RIS units configuration:
	\begin{equation}
	    \eta = \dfrac{\frac{1}{K}\Big(K_1+\sum_{k=1}^{K_2}u_{k}\Big)}{\frac{1}{|\mathcal{U}|}\sum_{k=1}^{K_2}  \Big(P_{k }^{\rm CS}u_k + P_{\rm RIS}N_ku_k\Big)},
	\end{equation}
	where $u_k$ is the indicator variable, if $u_k=1$, user $k$ will get service by HAPS-RIS, otherwise not if $u_k=0$. In the denominator, the \textcolor{black}{terms} $\sum_{k=1}^{K_2} P_{k }^{\rm CS}u_k$ and $\sum_{k=1}^{K_2}P_{\rm RIS}N_ku_k$ represent the total transmit power consumption at the \ac{CS} and the total power consumed by the RIS units for all supported UEs, respectively. $P_{\rm RIS}$ denotes the consumed power by each RIS unit for phase shifting, which is dependent on the RIS configuration technology and its resolution. Finally, $|\mathcal{U}|$ denotes the cardinality of the set, which constitutes the UEs supported by the CS via HAPS-RIS.
	\end{definition}

	In the following, we formulate the resource-efficient UEs maximization problem.
		It can be expressed as
		\begin{subequations}
			\label{eq:RIS-efficiency}
			\begin{align}
			\label{eq:P4-2}
			&\max _{u_{k},\mathrm{\Phi}_{k},N_{k},P_{k }^{\rm CS}} \eta\\
			\label{p4:c1} & \text { s.t. } L \leq L_{\rm max}, \\
			\label{p1:c1} &\quad R_{k}  \geq u_{k} R_{\rm th}, \enspace \forall k=1,2, \ldots, K_2, \\
			\label{p1:c2} &\quad  \sum_{k=1}^{K_2}  u_{k}N_{k}  \leq  N_{\rm max}, \\
			\label{p1:c3} &\quad 
			\textcolor{black}{\phi_{i,k} \in \{0, \Delta \phi, 2\Delta \phi, \ldots, (2^b-1)\Delta \phi\},
			\enspace  \forall i=1,2, \ldots, N_{k},}\notag\\
			&\qquad \qquad \qquad \qquad \qquad \qquad \qquad \textcolor{black}{\forall k = 1,2, \ldots, K_2,} \\
			\label{p1:c4} &\quad\sum_{k=1}^{K_2} u_{k}P_{k }^{\rm CS}  \leq P_{\rm max}^{\rm CS},
			\\
			\label{p1:c5} &\quad  \textcolor{black}{N_{k} \in \{0,1,\ldots, N_{k, \rm max}\}, \text{ \enspace $\forall k=1,2, \ldots, K_2,$}}\\
			\label{p1:c6} &\quad  0 \leq P_{k }^{\rm CS}  \leq u_{k} P_{k, \rm max}^{\rm CS} ,\enspace \forall k=1,2, \ldots, K_2,\\
			\label{p1:c7} &\quad u_{k} \in \{0,1\},
			\end{align}
		\end{subequations}
		where 
		$P^{\rm CS}_{\rm max}$ denotes the maximum available transmit power at the \ac{CS}.  $N_{k,\rm max}$ and $P_{k, \rm max}^{\rm CS}$ denote the maximum number of RIS units and the amount of power allocated to UE $k$, respectively. (\ref{p4:c1}) limits the  number of BSs in the area. Constraint (\ref{p1:c1}) guarantees that each selected UE satisfies the \textcolor{black}{minimum rate requirement of $R_{\rm th}$}. Constraint (\ref{p1:c2}) ensures the total number of RIS units allocated to the supported UEs does not exceed the maximum number of available RIS units, $N_{\rm max}$, which is limited by the HAPS size. \textcolor{black}{Constraint (\ref{p1:c3}) determines the discrete range of the adjustable phase shifts values for each RIS unit, where $\Delta\phi_{i,k}= 2\pi/2^b$ with $b$ as the number of bits used to uniformly quantize the phase shifts of each RIS element.} Constraint (\ref{p1:c4}) guarantees that the total allocated power by the \ac{CS} to the selected UEs is less than the maximum available power. Finally, constraints (\ref{p1:c5}) and (\ref{p1:c6}) ensure \textcolor{black}{fair allocation of both RIS units and \ac{CS} power\textcolor{black}{, respectively,} to each UE.} 

	\section{Proposed Solution} \label{Sec:Solution}

	Since problem (\ref{eq:RIS-efficiency}) is an \ac{MINLP}, it is hard to solve \textcolor{black}{jointly with the involved variables in an optimal manner with lower complexity. Therefore, we decouple the problem into two subproblems and }develop a low-complexity algorithm to solve it suboptimally. The main step of the proposed algorithm involves solving \eqref{eq:RIS-efficiency} in two stages. 
		
		In the first \textcolor{black}{subproblem}, we maximize the numerator of (\ref{eq:P4-2})  by maximizing the number of UEs, which can establish direct connection with TNs (i.e., by setting $L=L_{\rm max}$\footnote{\textcolor{black}{This value is dependent on the communication frequency, and the statistics of the users density and their rate demands. It is also determined by operator's expenditure analysis.}}), and then finding the set of maximum feasible UEs $ \in \mathcal{K}_2$, which can be supported by the CS via HAPS-RIS.  
		To this end, we first sort the channel gains of all UEs. Secondly, under the assumptions of equal power allocation to each UE and perfect reflection of each RIS unit, we allocate the minimum required number of RIS units to each UE as follows:
	\begin{equation}\label{eq:N_k}
	N_{k} = \Biggl\lceil \sqrt{\frac{N_0 B_{\rm UE} (2^{\frac{R_{\rm min}}{B_{\rm UE}}}-1)}{P_{k }^{\rm CS} \left| h_{k}  \right|^2}}\Biggr\rceil.
	\end{equation}
	

    The initial RIS units allocation starts with UEs with the best channel conditions until all RIS units are utilized. As a result, a set $\mathcal{U}$ with the largest number of feasible UEs is determined. 
	
    In the second \textcolor{black}{subproblem}, 
		the denominator of  (\ref{eq:P4-2}) is minimized by optimally allocating the \ac{CS} power and RIS units to each UE belongs to set $\mathcal{U}$. This is accomplished by solving the following optimization problem:
		\begin{subequations}\label{eq:min_N_general}
			\begin{align}
			\label{eq:min_N}
			&\min _{\mathrm{\Phi}_{k},N_{k},P_{k }^{\rm CS}} \sum_{k=1}^{|\mathcal{U}|}  P_{k }^{\rm CS} + P_{\rm RIS}N_k \\
			\label{p3:c1} &\text { s.t. } \textcolor{black}{R_{k} \geq R_{\rm th}}, \enspace \forall k=1,2, \ldots, |\mathcal{U}|, \\
			\label{p1_throu:c3} &\quad  \sum_{k=1}^{K_2}  N_{k} \leq N_{\rm max}, \\
			\label{p1_throu:c4} &\quad 
		\textcolor{black}{\phi_{i,k} \in \{0, \Delta \phi, 2\Delta \phi, \ldots, (2^b-1)\Delta \phi\},
			\enspace  \forall i=1,2, \ldots, N_{k},}\notag\\
			&\qquad \qquad \qquad  \qquad \qquad \qquad \qquad \textcolor{black}{\forall k = 1,2, \ldots, K_2,} \\
			\label{p1_throu:c5} &\quad\sum_{k=1}^{K_2} P_{k }^{\rm CS} \leq P_{\rm max}^{\rm CS},
			\\
			\label{p1_throu:c6} &\quad 
			\textcolor{black}{N_{k} \in \{0,1,\ldots, N_{k, \rm max}\},  \enspace \forall k=1,2, \ldots, |\mathcal{U}|,}\\
			\label{p1_throu:c7} &\quad 0 \leq P_{k }^{\rm CS} \leq P_{k, \rm max}^{\rm CS}, \enspace \forall k=1,2, \ldots, |\mathcal{U}|.
			\end{align}
		\end{subequations}
		
		Without loss of generality, problem \eqref{eq:min_N_general} can be re-written as
		\begin{subequations}\label{eq:min_N_modified}
			\begin{align}
			\label{eq:min_N_modified_a}
			&\min _{\mathrm{\Phi}_{k},N_{k},P_{k }^{\rm CS}} \sum_{k=1}^{|\mathcal{U}|}  P_{k }^{\rm CS} + P_{\rm RIS}N_k \\
			\label{p3:c1_modified} &\text { s.t. }  \frac{1}{\gamma_{k}} \leq \frac{1}{\gamma_{\rm min}}, \enspace \forall k=1,2, \ldots, |\mathcal{U}|, \\
			\label{p1_throu:c3_mod} &\quad \quad \eqref{p1_throu:c3} - \eqref{p1_throu:c7}.  
			\end{align}
		\end{subequations}

	\textcolor{black}{Due to the involvement of integer and discrete variables $N_{k}$ and $\phi_{i,k}$, respectively, problem \eqref{eq:min_N_modified} is a challenging problem and  is difficult to solve optimally in polynomial time. To solve it efficiently, we relax $N_{k}$ and $\phi_{i,k}$ to be  continuous variables. 
	 After this relaxation, the objective and the constraints of \eqref{eq:min_N_modified} become posynomials\footnote{\textcolor{black}{A posynomial function is a special function refers to a sum of positive monomials\cite{boyd2007tutorial}.}}, which can be solved optimally using geometric programming (GP) technique \cite{boyd2007tutorial}. At the solution of the relaxed problem, the final solution to \eqref{eq:min_N_modified} is obtained by approximation as $N_{k} \approx \lceil N^*_{k} \rceil$. The pseudo-code of the proposed two-stage algorithm is described in Algorithm~\ref{Algo1}. Note that in practice $\phi_{i,k}$ should be selected from a set of discrete phase shifts. The cardinality of the set is based on the resolution of the phase shift. The selection optimization of the discrete phase shifts is still an open problem. We investigated this problem in \cite{rivera2022optimization} and showed that with our proposed algorithm, the performance gap between continuous phase shifts and a set of two discrete phase shifts is about 30\%. The impact of relaxing $\phi_{i,k}$ and the performance gap can be significantly reduced with a set of large number of phase shifts.}

	\textcolor{black}{\subsubsection*{Complexity Analysis} The iteration-complexity of Algorithm~\ref{Algo1} is the same as of the standard barrier-based interior-point method with accuracy error $\epsilon$, and is in the order of $\mathcal{O}(2.5\sqrt{|\mathcal{U}|+1}\ln\frac{6(|\mathcal{U}|+1)\Delta}{\epsilon})$, where the term $\Delta$ is associated to a perturbation in the feasible set \cite{9709573}. However, the complexity of solving the original formulated problem optimally without relaxations and without decoupling them into two subproblems is computationally prohibited as it requires searching all possible combinations of RIS unit allocation and phase shifts configurations, and can be given as $\mathcal{O}\big((2^b)^{N_{\rm max}}\sum_{k=1}^{K_2}\prod_{i=1}^{N_{k, \rm max}}\binom{N_{\rm max}}{i}\big)$.}

	\begin{algorithm}[h]
		\caption{Efficient maximization of connected UEs}
		\label{Algo1}
		\begin{algorithmic}[1]
		    \State  Set $L= L_{\rm max}$ and obtained set $\mathcal{K}_2$.
			\State \textbf{Input:} $h_{k}, k\in \{1, 2, \ldots, K_2\}$. 
			\State  Sort $K_2$ UEs in descending order based on their channel gains $K_{S2}\leftarrow K_2$
			\For{$k = 1 \dots K_{S2}$} \;  
			\While{$\sum_{k=1}^{K_{S2}}  N_{k}  \leq N_{\rm max}$} 
			\State $P_{k }^{\rm CS}=P_{\rm max}^{\rm CS}/\textcolor{black}{K_{S2}}$
			\State Obtain initial $N_{k}$ from (\ref{eq:N_k}).
			\EndWhile
			\EndFor 
			\State \textbf{Stage-1 Output:} Selected UEs $\mathcal{U}$.
			\State Solve optimally 
			\eqref{eq:min_N_modified}
			for the set $\mathcal{U}$.  
			\State \textbf{Stage-2 Output:} $P_{k }^{\text{CS}*}$ and $N^*_{k}$ ($\forall k=1,\ldots,|\mathcal{U}|$).
		\end{algorithmic}
	\end{algorithm}
	\section{Numerical Results and Discussion} \label{Sec:Results}
	
	
		In this section, we present and discuss the performance of the  proposed  algorithm. \textcolor{black}{For the purpose of comparison}, we employ a benchmark approach, which first allocates equal power to all UEs and then selects the UEs with the highest channel gains to be served first with the minimum number of RIS units until the QoS requirement is satisfied.

		In the simulation setup, we consider an urban square area \textcolor{black}{of dimension} 10 km by 10 km, \textcolor{black}{with $L_{\rm max}=4$} terrestrial BSs serving $K=100$ uniformly randomly distributed UEs with a minimum separation distance of 100 m among them. We assume that the terrestrial BSs are optimally placed in the considered region. The channel gains between all the UEs and the terrestrial BSs are obtained by \textcolor{black}{adopting} the 3GPP standards \cite{3gpp38901study}. The carrier frequency is set to $f_c=2$ GHz with shadowing standard deviation $\sigma=8$ dB. Unless stated otherwise, the minimum rate for a direct connection between a UE and a BS is $R_{\rm th} = 2$ Mb/s. The maximum available bandwidth to each BS is $B_{\rm BS}=50$ MHz, and subcarrier bandwidth for each UE is set to $B_{\rm UE}=2$ MHz. Accordingly, a UE will be connected to a terrestrial BS that provides the highest data rate, and the collection of such UEs forms the set $\mathcal{K}_1$. Consequently, the collection of stranded UEs forms the other set $\mathcal{K}_2$.

\begin{figure}
		\centering
		\includegraphics[scale=0.6]{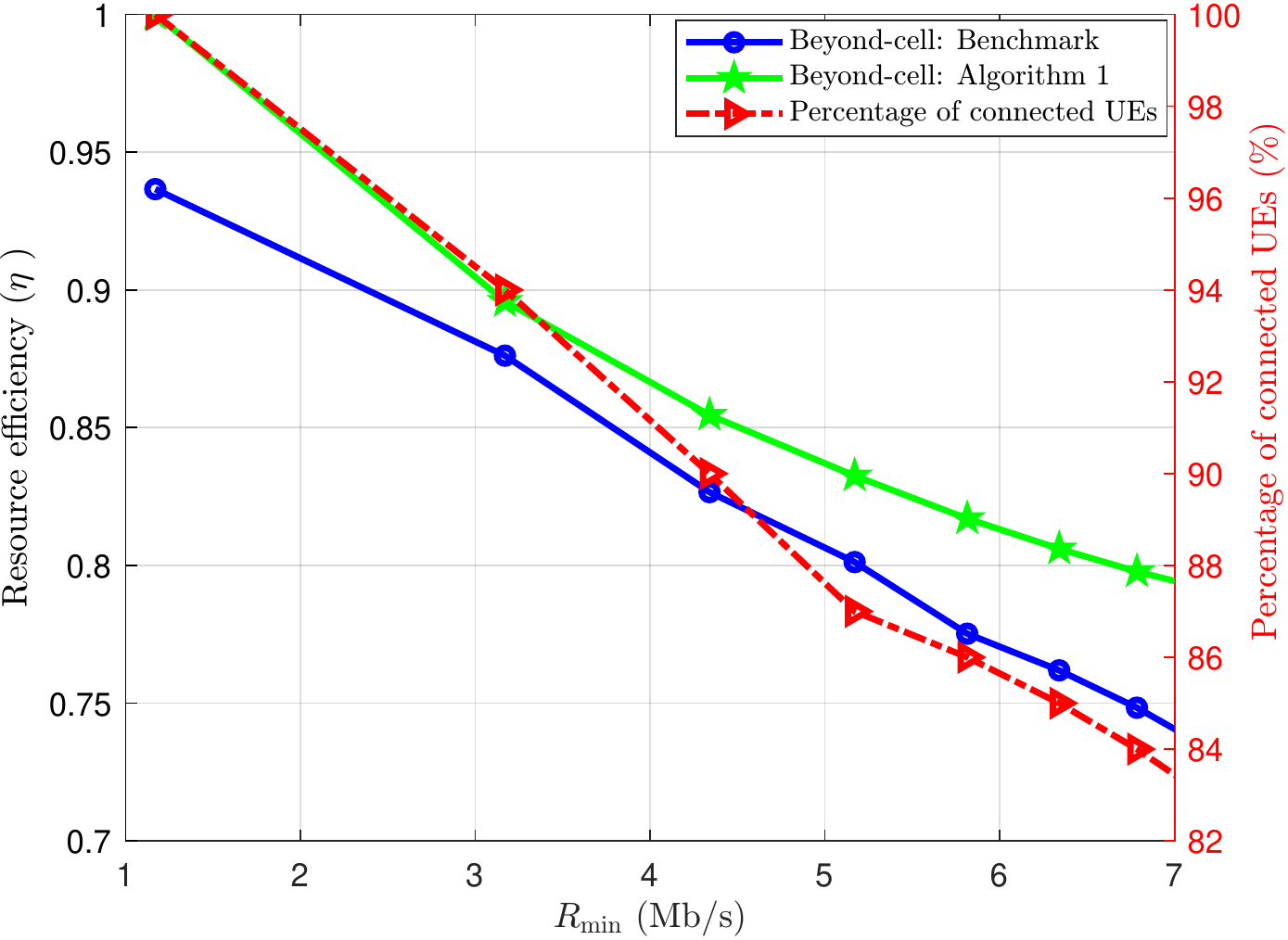}
		\caption{Resource-efficiency performance of connected UEs  for different $R_{\rm min}$.}
		\label{fig:max_UES_minPCS} 
	\end{figure}

		On the other hand, the effective channel gains from the \ac{CS} to all the UEs in set  $\mathcal{K}_2$ through HAPS-RIS are obtained using the standardized 3GPP channel model between a HAPS and terrestrial nodes in urban environments \cite{3gpp2017Technical,alfattani2021link}. 
		 This model considers dry air atmospheric attenuation, and corresponding atmosphere parameters are selected based on the mean annual global reference atmosphere \cite{itu1999p}. Further, we assume each UE has 0 dB antenna gain, and noise power spectral density $N_0 = -174$ dBm/Hz. We further set  $P_{\rm max}^{\rm CS} =$  33 dBm, and $G^{\rm CS}=$ 43.2 dBm \cite{3gpp2017Technical} in all of the simulations, unless stated otherwise. 
		The values of other parameters are as follow: $P_{\rm RIS} =7.8$ mW\cite{huang2019reconfigurable}, $P_{k, \rm max}^{\rm CS}=30$ dBm, and $N_{k, \rm max}=50,000$ units.
			{\color{black}

		\subsection{Resource-efficiency maximization}
        Fig. \ref{fig:max_UES_minPCS} plots the normalized RE obtained using Algorithm~\ref{Algo1} (on the left-hand side y-axis) and the percentage of connected UEs ( on the right-hand side y-axis) versus different values of minimum rate requirement $R_{\rm min}$. It also compares the performance of Algorithm~\ref{Algo1} with the benchmark approach. The maximum number of RIS units mounted on HAPS is set to $N_{\rm max}=220,000$ units. It can be observed that as the QoS (represented by $R_{\rm min}$) increases, the percentage of connected UEs and the RE drops. However, this performance degradation is more significant in terms of the percentage of connected UEs than the RE. Furthermore, we observed that the RE obtained using Algorithm~\ref{Algo1} significantly outperforms the one obtained using the benchmark approach. This is due to the fact that Algorithm~\ref{Algo1} optimizes allocation of both power and RIS units to UEs.
	} 
	
	
		\subsection{Percentage of connected UEs}
		\begin{figure}
			\centering
			\includegraphics[scale=0.6]{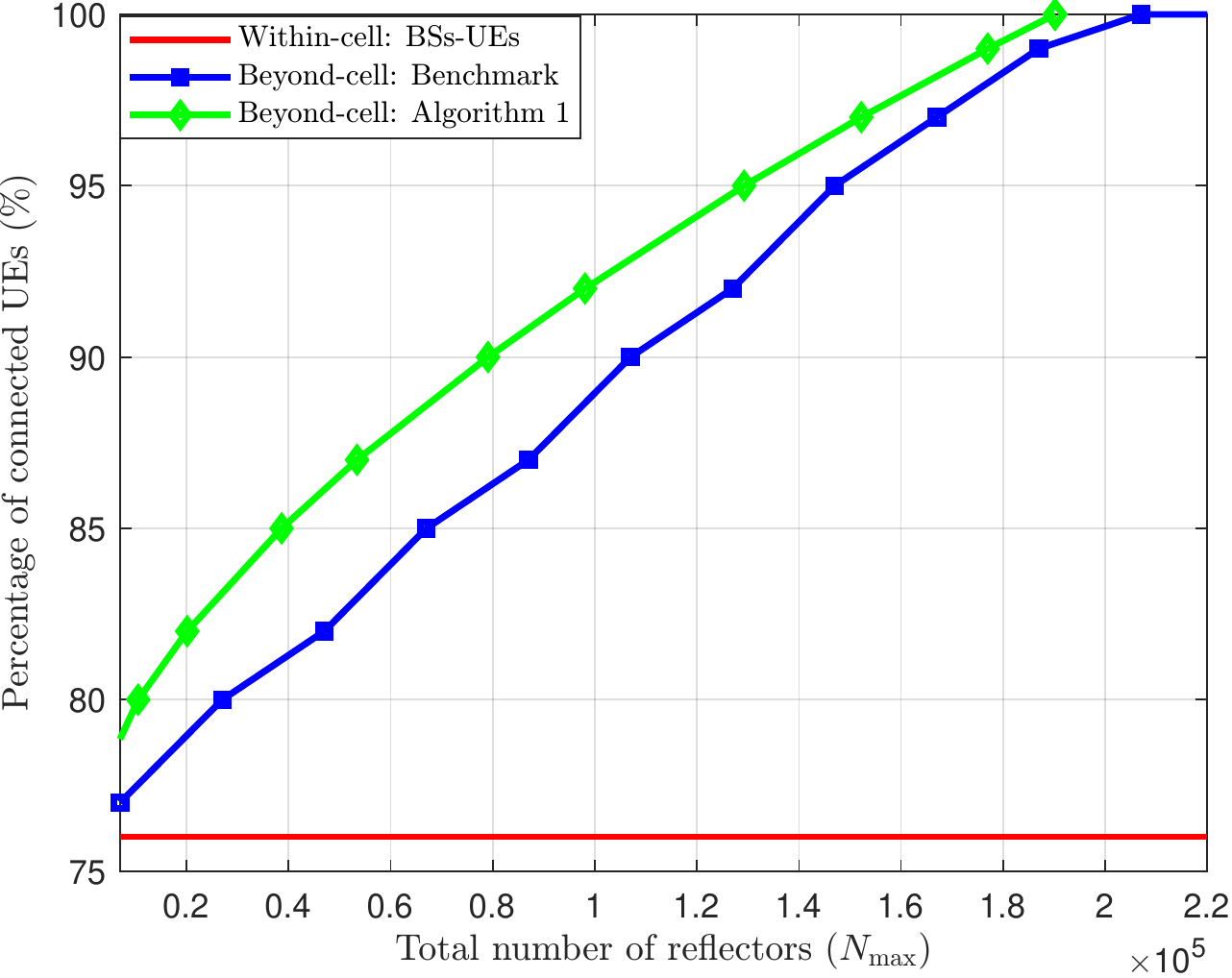}
			\caption{Maximizing connected UEs performance for different $N_{\rm max}$.}
			\label{fig:max_UES_algo} 
		\end{figure}

			\begin{figure}
			\centering
		\includegraphics[scale=0.6]{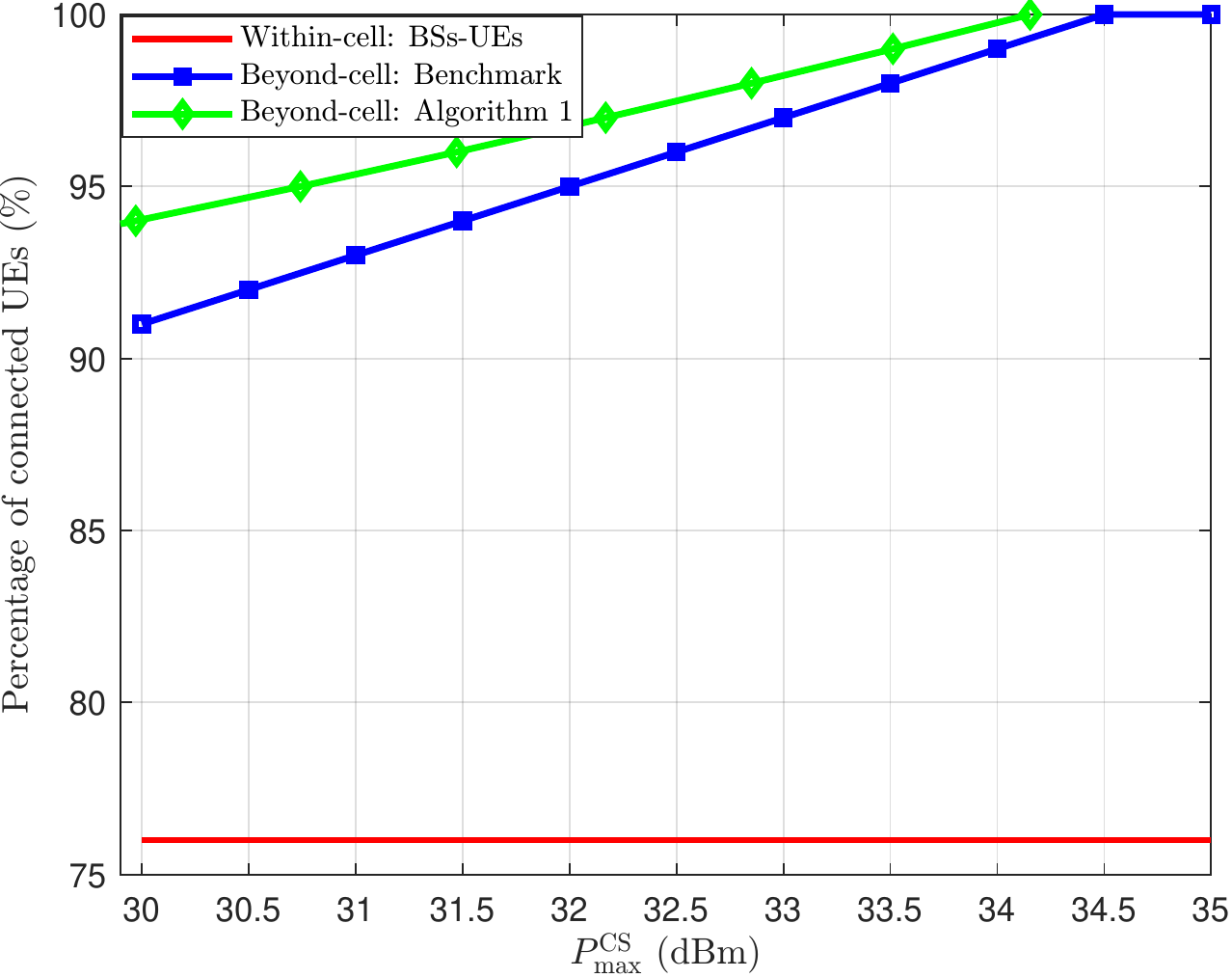}
			\caption{Maximizing connected UEs performance for different $P_{\rm max}^{\rm CS}$.}
		\label{fig:max_UES_algo_diff_PCS} 
		\end{figure}
		Figs. \ref{fig:max_UES_algo} and \ref{fig:max_UES_algo_diff_PCS} plot and compare the percentage of connected UEs obtained through Algorithm~\ref{Algo1}, the benchmark approach, and \textit{within-cell} communication approach for different values of the number of RIS units $N_{\rm max}$  available at the HAPS, and maximum power $P_{\rm max}^{\rm CS}$ available at the \ac{CS}, respectively. 

In Fig. \ref{fig:max_UES_algo}, to study the impact of $N_{\rm max}$, we consider only the second term (i.e., $P_{\rm RIS}N_k$) of the objective function in (\ref{eq:min_N_modified_a}). The selected range of $N_{\rm max}$ is set between 10,000 and 220,000 units. This range corresponds to a total RIS area between  9 $\rm m^2$ and 198 $\rm m^2$ at carrier frequency of 2 GHz \footnote{This represents a limited area on a typical HAPS surface, as the length of an airship is between	100-200 m, whereas aerodynamic HAPS  have wingspans between 35 m and 80 m. The size of each RIS unit is $(0.2\lambda)^2$ \cite{kurt2021vision}.}. In Fig. \ref{fig:max_UES_algo_diff_PCS}, only the first term of (\ref{eq:min_N_modified_a}) (i.e., $P_{k}^{\rm CS}$) is considered to study the effect of $P_{\rm max}^{\rm CS}$ on the percentage of connected UEs.  The maximum transmit power of CS is set to vary between 30 dBm and 35 dBm, and $N_{\rm max}$ is set to 150,000 units. It can be observed from the figures that the percentage of connected UEs increases with the maximum power of the \ac{CS} $P_{\rm max}^{\rm CS}$, and the maximum number of RIS units $N_{\rm max}$  (or the size of HAPS). These behaviours are intuitive as making \textcolor{black}{ more system resources ($P_{\rm max}^{\rm CS}$ and $N_{\rm max}$) available allows} more number of stranded users to be served by the \ac{CS} via HAPS-RIS.  

Figs. \ref{fig:max_UES_algo} and \ref{fig:max_UES_algo_diff_PCS} also show that the performance of the proposed approach is 1-3 \% higher than the
benchmark approach. Moreover, $P_{\rm max}^{\rm CS}$ has more significant impact than $N_{\rm max}$ on the percentage of connected UEs. By increasing $P_{\rm max}^{\rm CS}$ by 2 dB, the percentage of connected UEs increases about 4\%, whereas  doubling of the RIS size is needed to achieve the same increase in the percentage of connected UEs.

Furthermore, it can be observed from the figures that 76\%  UEs are served through \textit{within-cell} communication approach, and the \ac{CS} supports the remaining UEs via HAPS-RIS. Hence, \textit{beyond-cell} communications via HAPS-RIS is able to complement  TNs.

	\section{Conclusion}\label{Sec:conclstion}
	\textcolor{black}{In this letter, we investigated the resources optimization of \textit{beyond-cell} communications that use HAPS-RIS technology to complement TNs by supporting  unserved UEs. 
    In particular, given the limitations of the CS power and HAPS-RIS size, it might not be feasible to support all unserved UEs. 
    Therefore, we formulated
    a novel resource-efficient optimization problem that
    simultaneously maximizes the percentage of connected UEs using minimal \ac{CS} power and RIS units.
     The results show the capability of the \textit{beyond-cell} communications approach to support a larger number of UEs. Further, the results show the superiority of the proposed solutions over the benchmark approach and demonstrate the impact of the HAPS size and the QoS requirement on the percentage of connected UEs and the efficiency of the system.  } 

	\bibliographystyle{IEEEtran}
	\bibliography{IEEEabrv,final_version_WCL}
\end{document}